\documentclass[epj]{svjour}
\usepackage{latexsym}
\usepackage{graphics}
\usepackage{amssymb}

\newcommand{\be}[1]{\begin{equation}\label{#1}}
\newcommand{\ee}{\end{equation}}
\newcommand{\ba}[1]{\begin{eqnarray}\label{#1}}
\newcommand{\ea}{\end{eqnarray}}
\newcommand{\rf}[1]{(\ref{#1})}
\newcommand{\nn}{\nonumber}

\newcommand{\const}{\mbox{\rm const}\,}

\usepackage{bm}

\begin{document}

\title{Gravitational potentials and forces in the Lattice Universe: a slab}

\author{Maxim Eingorn\inst{1} \and Niah O'Briant\inst{1} \and Katie Arzu\inst{1} \and Maxim Brilenkov\inst{2} \and Alexander Zhuk\inst{3}}

\institute{Department of Mathematics and Physics, North Carolina Central University, \\ 1801 Fayetteville St., Durham, North Carolina 27707, U.S.A. \email{maxim.eingorn@gmail.com}
	\and Institute of Theoretical Astrophysics, University of Oslo, Postboks 1029 Blindern,
	0315, Oslo, Norway
	\and Astronomical Observatory, Odessa I.I. Mechnikov National University, \\ Dvoryanskaya St. 2, Odessa 65082, Ukraine}

\date{%Received: date / Revised version: date
}

\abstract{We study the effect of the slab topology $T\times R\times  R$ of the Universe on the form of gravitational potentials and forces created by point-like masses. We obtain two alternative forms of solutions: one is based on the Fourier series expansion of the delta function using the periodical property along the toroidal dimension, and another one is derived by direct summation of solutions of the Helmholtz equation for the source particle and all its images. The latter one takes the form of the sum of Yukawa-type potentials. We demonstrate that for the present Universe the latter solution is preferable for numerical calculations since it requires less terms of the series to achieve the necessary precision.
\PACS{
     {04.25.Nx}{Post-Newtonian approximation;
     	perturbation theory; related
     	approximations}   \and
      {98.80.Jk}{Mathematical and relativistic
      	aspects of cosmology}
} % end of PACS codes
} %end of abstract

\maketitle

\section{Introduction}\label{intro}

The question of spatial topology of the Universe belongs to a class of fundamental open questions of cosmology and theoretical physics. What is the shape of the world we live in? Is the Universe finite or infinite? Is its spatial curvature positive, negative or exactly zero? What is the role of topology in the very early epoch (on the quantum gravity arena) as well as for the subsequent large scale structure formation? Since topology is not dictated by general relativity, there is no theoretical hint of whether space is simply connected (as assumed within the concordance cosmological model) or multiply connected. In the latter case the Universe volume can be finite even when the spatial curvature is negative or vanishing \cite{36}. If the Universe volume is much larger than the observable one, the finiteness of the world does not become apparent in the current data. However, if the volume is not too large, it is reasonable to search for observable imprints of its shape \cite{37}. In multiply connected space, a photon emitted by a source can travel many times through the volume resulting in multiple images of the source \cite{43,44}. Typical representatives of multiply connected spaces are spaces with toroidal topology in one or several (maximum three) spatial directions: a slab $T\times R\times R$, an equal-sided chimney $T\times T\times R$ and a three-torus $T\times T\times T$. 

The possible imprints (mainly on the CMB data) of the shape of the Universe are carefully studied in literature \cite{38,39,40,41,42}. In particular, it is tempting to interpret the CMB anomalies observed at large angular scales (such as the quadrupole moment suppression as well as the quadrupole and octopole alignment) as topological manifestations \cite{47,48}. In the present paper, we consider the topology of the Universe in the form of a slab. In this case we have one finite special dimension. It is very interesting whether this direction can be interpreted as a preferred axis of the quadrupole and octopole alignment, or a so-called ``axis of evil'' \cite{49} (see also \cite{51} for other indications of its existence). 

According to Planck 2013 results \cite{36} regarding the search for conjectural topological signatures in the observational data on the CMB radiation, the following restriction is imposed on the radius $R_i$ of the largest sphere, which can be inscribed in the topological domain: in the case of the flat Universe with the slab topology $R_i > 0.50 \chi_{rec}$. Planck 2015 results \cite{42} make the above-mentioned restriction tougher: $R_i > 0.56 \chi_{rec}$. 
Here  $\chi_{rec}$  represents the distance to the recombination surface, which is of the order of the particle horizon, {\it i.e.} $\sim 14$ Gpc. Earlier bounds on the Universe size, based on the thorough analysis of 7-year and 9-year WMAP temperature maps, can be found in \cite{49,52}. A lower bound on the size of the fundamental topological domain in the case of the flat Universe, based on the 7-year WMAP data, is $d=2 R_{\mathcal{LSS}}\cos(\alpha_{min}) \simeq 27.9$Gpc \cite{48}, where $R_{\mathcal{LSS}}$ is the distance to the last scattering surface ({\it i.e.} the recombination surface).

In the present paper, we study gravitational properties of the Universe with the slab topology $T\times R\times R$. More precisely, we investigate the effect of such topology on the form of the gravitational potential and force. It is  well known that in the Newtonian limit the gravitational potential is defined by the Poisson equation. In the case of cosmology, the  matter density fluctuations are the sources of this potential \cite{Peebles}. For toroidal types of topology, this equation was investigated in \cite{topology1}. It was shown that there is no way to get any physically reasonable and nontrivial solution of this equation in the case of the slab topology. However, if we take into account relativistic effects and derive the equation for the gravitational potential from the perturbed Einstein equations, then we find that the resulting equation has the form of the Helmholtz equation rather than the Poisson one \cite{Eingorn1,Claus1,Claus2}. As we show in the present paper, this changes the situation drastically  and for the considered slab topology we obtain the physically reasonable solutions. Moreover, to achieve it, we do not make any artificial assumptions about the distribution of gravitating masses. We obtain two alternative forms of the gravitational potentials and forces and demonstrate that the solutions in the form of the sum of Yukawa potentials are preferable in the present Universe for numerical calculations.
   
The paper is structured as follows. In sect.~\ref{sec:2}, the basic equations are presented and two alternative forms of solutions for the  gravitational potential are obtained. They are compared from the point of numerical calculations in sect.~\ref{sec:3}. The corresponding alternative expressions for the gravitational force are derived in sect.~\ref{sec:4}. These expressions are also compared from the point of numerical calculations. A brief summary of the main results is given in concluding sect.~\ref{sec:5}.

\section{Basic equations and alternative solutions}\label{sec:2}

\setcounter{equation}{0}

If we take into account relativistic effects in the framework of the conventional $\Lambda$CDM cosmological model, then the gravitational potential satisfies the following equation \cite{Eingorn1,Claus1,Claus2}:
%%%%%%%
\be{2.1}
\Delta\Phi_0 - \frac{3\kappa\overline{\rho} c^2}{2a}{\Phi}_0=\frac{\kappa c^2}{2a}\left(\rho -\bar{\rho}\right)\, ,
\ee
%%%%%%%
where $\kappa\equiv 8\pi G_N/c^4$, $G_N$ is the Newtonian gravitational constant, $c$ is the speed of light,  $a$ is the scale factor and $\Delta$ is the Laplace operator in comoving coordinates. We consider matter in the form of discrete point-like gravitating masses $m_n$ with comoving mass density
%%%%%%
\be{2.2}
\rho = \sum_n m_n\delta(\mathbf{r}-\mathbf{r}_n)\, .
\ee
%%%%%%
The averaged comoving mass density is constant: $\bar\rho =\const$. 
The subscript $0$ for $\Phi$ indicates that peculiar velocities were not taken into account in eq.~\rf{2.1} (see also \cite{cosmlaw}).

It can be easily seen that the shifted gravitational potential 
%%%%%
\be{2.3}
\widehat{\Phi}_0\equiv \Phi_0 -\frac13
\ee
%%%%%
satisfies the equation
%%%%%%
\be{2.4}
\Delta\widehat{\Phi}_0 - \frac{a^2}{\lambda^2} \widehat{\Phi}_0=\frac{\kappa c^2}{2a}\rho\, , 
\ee
%%%%%%
where we introduce the screening length \cite{Eingorn1}
%%%%%%%
\be{2.5}
\lambda \equiv \left(\frac{3\kappa\overline{\rho} c^2}{2a^3}\right)^{-1/2}\, .
\ee
%%%%%%%
Eq.~\rf{2.4} now allows us to apply the superposition principle to get its solution. First, we can find a solution for a single particle, let it be a particle $m$ in the center of the Cartesian coordinates, and then we can write a solution for the full system of particles.

In the case of the slab topology $T\times R \times R$, each gravitating mass $m_n$ has its counterparts shifted by a distance multiple of the torus period $l$ (for instance, along the $z$-axis in the Cartesian coordinates). Thus, for a selected particle $m$ we should take into account all its counterparts. To this end, and also using the topology of the system, we can present the delta function $\delta(z)$ in the form
%%%%%%
\be{1} \delta(z)=\frac{1}{l}\sum_{k=-\infty}^{+\infty}\cos\left(\frac{2\pi k}{l}z\right)\, .\ee
%%%%%%

Therefore, for a single point-like mass $m$ in the center of the coordinate system we have
%%%%%%
\ba{2}
&&\Delta\widehat{\Phi}_0-\frac{a^2}{\lambda^2}\widehat{\Phi}_0=\frac{\kappa c^2}{2a}\frac{m}{l}\sum_{k=-\infty}^{+\infty} \cos\left(\frac{2\pi
	k}{l}z\right)\delta\left(x\right)\delta\left(y\right)\, . \ea
%%%%%%

It is natural to seek for the solution in the following form:
%%%%%%
\be{3} \widehat{\Phi}_0=\sum_{k=-\infty}^{+\infty} C_{k}(x,y)\cos\left(\frac{2\pi k}{l}z\right)\, .\ee
%%%%%%
Substitution of this expression into eq.~\rf{2} gives
%%%%%%%
\be{6} \sum_{k=-\infty}^{+\infty}\left[\frac{\partial^2}{\partial x^2}C_{k}(x,y)+\frac{\partial^2}{\partial y^2}C_{k}(x,y)-
\left(\frac{4\pi^2k^2}{l^2}+\frac{a^2}{\lambda^2}\right)C_{k}(x,y)- \frac{\kappa
	c^2}{2a}\frac{m}{l}\delta\left(x\right)\delta\left(y\right)\right]\cos\left(\frac{2\pi k}{l}z\right)=0 \, .\ee
%%%%%%%%

It is convenient to use the polar coordinates:
%%%%%
\be{7} x=\xi\cos\phi, \quad y=\xi\sin\phi\, .\ee
%%%%%%%
Obviously, due to the symmetry of the model, coefficients $C_k$ depend only on the polar radius $\xi$, and for $\xi>0$ satisfy the equation
%%%%%
\be{10} \xi\frac{d^2 C_{k}}{d\xi^2}+\frac{d C_{k}}{d\xi}-\left(\frac{4\pi^2k^2}{l^2}+\frac{a^2}{\lambda^2}\right)\xi C_{k}(\xi)=0\, .\ee
%%%%%%
The general solution of this equation is superposition of the modified Bessel functions:
%%%%%%
\be{12} C_k(\xi)=AI_0\left(\sqrt{b}\xi\right)+BK_0\left(\sqrt{b}\xi\right),\quad b\equiv \frac{4\pi^2k^2}{l^2}+\frac{a^2}{\lambda^2}\, ,
\ee
%%%%%%
where $A$ and $B$ are the constants of integration.
Hence, omitting the growing mode $I_0\left(\sqrt{b}\xi\right)$, we get
%%%%%%
\be{13} C_{k}(\xi)=BK_0\left(\sqrt{\frac{4\pi^2k^2}{l^2}+\frac{a^2}{\lambda^2}}\xi\right)\, .\ee
%%%%%%
To define the constant $B$, we take into account that at small $\xi$ this function should satisfy the two-dimensional Poisson equation with a source proportional to $\delta(x)\delta(y)$. When $\xi\rightarrow0$, we have $C_k(\xi)\rightarrow-B\ln\xi$. On the other hand, $\triangle(\ln\xi)=2\pi\delta(x)\delta(y)$. So,
%%%%%%
\be{14} -2\pi B=\frac{\kappa c^2}{2a}\frac{m}{l},\quad B=-\frac{\kappa c^2}{4\pi a}\frac{m}{l}\, .\ee
%%%%%%
Hence,
%%%%%%
\be{15} C_{k}(\xi)=-\frac{\kappa c^2}{4\pi a}\frac{m}{l}K_0\left(\sqrt{\frac{4\pi^2k^2}{l^2}+\frac{a^2}{\lambda^2}}\xi\right)\, \ee
%%%%%%
and
%%%%%%%
\be{16} \widehat{\Phi}_0=-\frac{\kappa c^2}{4\pi a}\frac{m}{l}\sum_{k=-\infty}^{+\infty} K_0\left(\sqrt{\frac{4\pi^2k^2}{l^2}+\frac{a^2}{\lambda^2}}\xi\right)\cos\left(\frac{2\pi k}{l}z\right)\, .\ee
%%%%%%
Obviously, for a system of gravitating masses with arbitrary positions we have
%%%%%%
\be{17} \Phi_0=\frac{1}{3}-\frac{\kappa c^2}{4\pi a}\frac{1}{l}\sum_n m_n\left\{\sum_{k=-\infty}^{+\infty}
K_0\left(\sqrt{\frac{4\pi^2k^2}{l^2}+\frac{a^2}{\lambda^2}}\left|\bm{\xi}-\bm{\xi}_n\right|\right)\cos\left[\frac{2\pi
	k}{l}\left(z-z_n\right)\right]\right\}\, .\ee
%%%%%%

It is not difficult to demonstrate that the formula \rf{16} has the correct Newtonian limit in the vicinity of the considered gravitating mass. At such small distances all directions should be considered on an equal footing and summation is replaced by integration: 
%%%%%
\ba{22} \widehat{\Phi}_0\rightarrow-\frac{\kappa c^2m}{4\pi a}\int\limits_{-\infty}^{+\infty} K_0\left(\sqrt{4\pi^2\tilde k^2+\frac{a^2}{\lambda^2}}\xi\right)\cos\left(2\pi \tilde k z\right)d\tilde k\, , \quad \tilde k=\frac{k}{l}\, ,\  d\tilde k=\frac{dk}{l}\, .\ea
%%%%%%
We proceed with changing again the integration variable: $\tilde k=\tilde k_{\xi}/\xi$, $d\tilde k=d\tilde k_{\xi}/\xi$. Then, with the help of the formula 2.16.14(1) from \cite{Table}, we obtain
\ba{23new} \widehat{\Phi}_0&\rightarrow&-\frac{\kappa c^2m}{4\pi a}\frac{1}{\xi}\int\limits_{-\infty}^{+\infty} K_0\left(\sqrt{4\pi^2\tilde k_{\xi}^2+\frac{a^2}{\lambda^2}\xi^2}\right)\cos\left(2\pi \tilde k_{\xi}\frac{z}{\xi}\right)d\tilde k_{\xi}
\nn\\
&\rightarrow&-\frac{\kappa c^2m}{2\pi a}\frac{1}{\xi}\int\limits_{0}^{+\infty} K_0\left(2\pi \tilde k_{\xi}\right)\cos\left(2\pi \tilde k_{\xi}\frac{z}{\xi}\right)d\tilde k_{\xi}=-\frac{\kappa c^2m}{2\pi a}\frac{1}{\xi}\frac{1}{4\sqrt{1+z^2/\xi^2}}=-\frac{G_Nm}{c^2}\frac{1}{\sqrt{\Xi^2+Z^2}}\, ,\ea
where $\Xi=a\xi$ and $Z=az$ are the physical coordinates. This formula is exactly the Newtonian expression.

Now we want to demonstrate another important property of the gravitational potential $\Phi_0$. Since $\Phi_0$ is the linear fluctuation of the metric coefficients, its averaged value should be equal to zero \cite{Eingorn1} (see also argumentation in \cite{EBV}). Let us prove it. First, we rewrite Eq.~\rf{16} as
%%%%%%
\be{18} \widehat{\Phi}_0=-\frac{\kappa c^2}{4\pi a}\frac{m}{l}K_0\left(\sqrt{\frac{3\kappa\overline{\rho} c^2}{2a}}\xi\right)-\frac{\kappa c^2}{2\pi
	a}\frac{m}{l}\sum_{k=1}^{+\infty} K_0\left(\sqrt{\frac{4\pi^2k^2}{l^2}+\frac{a^2}{\lambda^2}}\xi\right)\cos\left(\frac{2\pi k}{l}z\right) \, .\ee
%%%%%%%
Therefore, 
%%%%%%%
\ba{19} &&\int\limits_{-\infty}^{+\infty}dx\int\limits_{-\infty}^{+\infty}dy\int\limits_{0}^{l}dz\widehat{\Phi}_0=-\frac{\kappa c^2}{4\pi
	a}m\int\limits_{-\infty}^{+\infty}dx\int\limits_{-\infty}^{+\infty}dy K_0\left(\sqrt{\frac{3\kappa\overline{\rho}
		c^2}{2a}\left(x^2+y^2\right)}\right)\nn\\
&=&-\frac{\kappa c^2}{4\pi a}m\cdot2\pi\int\limits_{0}^{+\infty}\xi d\xi K_0\left(\sqrt{\frac{3\kappa\overline{\rho} c^2}{2a}}\xi\right)=-\frac{\kappa c^2}{2
	a}m\left(\frac{3\kappa\overline{\rho} c^2}{2a}\right)^{-1}=-\frac{1}{3}\frac{m}{\overline\rho} \, ,\ea
%%%%%%%
where we have used the table integral 2.16.2(2) from \cite{Table}. 
Then, taking into account the relation \rf{2.3}, for the averaged value of the total gravitational potential $\Phi_0$ we have
%%%%%%
\be{20} \overline{\Phi_0}=\frac{1}{3}-\frac{1}{3}\frac{m}{\overline\rho}\cdot \frac{N}{L_xL_yl}=\frac{1}{3}-\frac{1}{3}=0,\quad \frac{mN}{L_xL_yl}=\overline\rho\, ,\ee
%%%%
where for simplicity we consider the case when all $N$ particles in the volume $V=L_x L_yl$ have the same mass $m$.

Above we have presented one of the possible ways to solve eq.~\rf{2.4}. However, since this is the Helmholtz equation, we can also solve it by direct summation over all counterpart contributions which have the form of Yukawa potentials:  
%%%%%%
\be{2.25} 
\widehat{\Phi}_{0}=-\frac{\kappa c^2}{8\pi a}\frac{m}{l}\sum_{k=-\infty}^{+\infty} \frac{l}{\sqrt{\xi^2+(z-kl)^2}}\exp\left(-\frac{a\sqrt{\xi^2+(z-kl)^2}}{\lambda}\right)\, .
\ee
%%%%%%%

As we already mentioned, eq.~\rf{2.4} with the corresponding solutions \rf{16} and \rf{2.25} does not take into account the peculiar velocities of gravitating masses. However, in \cite{MaxEzgi}, the authors argued the importance of such account. It was demonstrated that the peculiar velocities can be included back into consideration effectively in eq.~\rf{2.4} by the replacement of the screening length $\lambda $ with an effective screening length $\lambda_{\mathrm{eff}}$ (defined by the formula (41) in \cite{MaxEzgi}):   
%%%%%%
\be{28}\triangle\widehat\Phi-\frac{a^2}{\lambda_{\mathrm{eff}}^2}\widehat\Phi=\frac{\kappa c^2}{2a}\rho\, . 
\ee 
%%%%%%
In particular, at the matter-dominated stage  $\lambda_{\mathrm{eff}}=\sqrt{3/5}\lambda$. Hence, instead of \rf{16}, \rf{17} we have, respectively,
\be{29} \widehat{\Phi}=-\frac{\kappa c^2}{4\pi a}\frac{m}{l}\sum_{k=-\infty}^{+\infty} K_0\left(\sqrt{\frac{4\pi^2k^2}{l^2}+\frac{a^2}{\lambda_{\mathrm{eff}}^2}}\, \xi\right)\cos\left(\frac{2\pi k}{l}z\right)\, ,\ee
\be{30} \Phi=\frac{1}{3}\left(\frac{\lambda_{\mathrm{eff}}}{\lambda}\right)^2-\frac{\kappa c^2}{4\pi a}\frac{1}{l}\sum_n m_n\left\{\sum_{k=-\infty}^{+\infty}
K_0\left(\sqrt{\frac{4\pi^2k^2}{l^2}+\frac{a^2}{\lambda_{\mathrm{eff}}^2}}\, \left|\bm{\xi}-\bm{\xi}_n\right|\right)\cos\left[\frac{2\pi k}{l}\left(z-z_n\right)\right]\right\}\, .\ee
The average value $\overline{\Phi}$ still equals $0$.

In the case of the solution in the form of \rf{2.25}, we should simply replace $\lambda$ with $\lambda_{\mathrm{eff}}$. To distinguish among the derived expressions for the gravitational potential, we introduce the following notation:
%In the present section we have derived two different forms of the solution
%of eq.~\rf{2.4}. They are eqs.~\rf{16} and \rf{2.25}.
%To distinguish them, we introduce the following notation:
%%%%%
\be{2.29}
\tilde{\Phi}_{\cos} \equiv \left(-\frac{\kappa c^2}{8\pi a}\frac{m}{l}\right)^{-1} \widehat{\Phi}_{\cos}=2\sum_{k=-\infty}^{+\infty} K_0\left(\sqrt{4\pi^2k^2+\frac{1}{\tilde{\lambda}_{\mathrm{eff}}^2}}\tilde\xi\right)\cos\left(2\pi k\tilde z\right)\,
\ee
%%%%%
and
%%%%%%
\be{2.30}
\tilde{\Phi}_{\exp}\equiv \left(-\frac{\kappa c^2}{8\pi a}\frac{m}{l}\right)^{-1}\widehat{\Phi}_{\exp}=\sum_{k=-\infty}^{+\infty} \frac{1}{\sqrt{\tilde\xi^2+(\tilde z-k)^2}}\exp\left(-\frac{\sqrt{\tilde\xi^2+(\tilde z-k)^2}}{\tilde\lambda_{\mathrm{eff}}}\right)\, ,
\ee
%%%%%%
where the rescaled quantities are:
%%%%%%%%
\be{2.31} 
z=\tilde zl,\quad \xi=\tilde\xi l,\quad \lambda_{\mathrm{eff}}=\tilde\lambda_{\mathrm{eff}}al\, .
\ee
%%%%%%
In what follows, we will explore the benefits of one formula over another from the point of view of numerical analysis.
%%%%%%%%%%%%%%%%%%%%%%%%%%%%%%%%%%%%%%%%
%%%%%%%%%%%%%%%%%%%%%%%%%%%%%%%%%%%%%%%%%%%
%%%%%%%%%%%%%%%%%%%%%%%%%%%%%%%%%%%%%%%%%%

\section{Gravitational potentials}\label{sec:3}

\setcounter{equation}{0}

The functions \rf{2.29} and \rf{2.30} represent the contribution to the gravitational potential (scalar perturbation), produced by a point-like mass $m$ located at $z=0$, $\xi=0$, and its images located at $z=\pm l,\pm 2l,\ldots$, $\xi=0$, where $l$ is the (comoving) period of the torus, $z$ is the coordinate along the toroidal dimension, and $\xi$ is the distance from the $z$-axis. 

For numerical calculations, we have to cut off the infinite series \rf{2.29} and \rf{2.30} at some number $n$. This number of terms depends on the precision with which we want to calculate these expressions. Obviously, the fewer the number of terms needed for this, the better the corresponding formula is suitable for computation. We want to compare the formulas \rf{2.29} and \rf{2.30} from this point of view. The first conclusion, which follows from eq.~\rf{2.29}, is that this expression for any $n$ does not work for points with $\xi=0$, since the modified Bessel function $K_0$ diverges at zero value of the argument. On the other hand, there is no such limitation for eq.~\rf{2.30}.

Let us introduce the total numbers of terms $n_{\cos}$ and $n_{\exp}$ which we would like to include in the computation, then it follows from \rf{2.29} and \rf{2.30} that
%%%%%%
\be{3.1} 
\left.\tilde{\Phi}_{\cos}\right|_{n_{\cos}}=2K_0\left(\frac{\tilde\xi}{\tilde\lambda_{\mathrm{eff}}}\right)+4\sum_{k=1}^{n_{\cos}-1} K_0\left(\sqrt{4\pi^2k^2+\frac{1}{\tilde\lambda_{\mathrm{eff}}^2}}\tilde\xi\right)\cos\left(2\pi k\tilde z\right)\, \ee
%%%%%%%
and
%%%%%%%
\ba{3.2} 
&{}&\left.\tilde{\Phi}_{\exp}\right|_{n_{\exp}}=\frac{1}{\sqrt{\tilde\xi^2+\tilde z^2}}\exp\left(-\frac{\sqrt{\tilde\xi^2+\tilde z^2}}{\tilde\lambda_{\mathrm{eff}}}\right)\nn\\
&+&
\sum_{k=1}^{n_{\exp}-1} \left[\frac{\exp\left(-\sqrt{\tilde\xi^2+(\tilde z+k)^2}/\tilde\lambda_{\mathrm{eff}}\right)}{\sqrt{\tilde\xi^2+(\tilde z+k)^2}}+\frac{\exp\left(-\sqrt{\tilde\xi^2+(\tilde z-k)^2}/\tilde\lambda_{\mathrm{eff}}\right)}{\sqrt{\tilde\xi^2+(\tilde z-k)^2}}\right]\, ,\ea
%%%%%%%
where we singled out the zero modes. The results of calculations with the help of Mathematica \cite{Math} are presented in Tables~\ref{results_table_1},~\ref{results_table_2}. 
The values of $n_{\exp}$ in these tables describe the number of terms in \rf{3.2}, required to achieve the four-digit accuracy of determining $\tilde{\Phi}$ (at the point of interest with some coordinates $z,\xi$). For all $n\geqslant n_{\exp}$ the four-digit value of $\left.\tilde{\Phi}_{\exp}\right|_n$ does not change.
% and defines so-called the "exact" value $\tilde\Phi$.
If eq.~\rf{3.1} is used instead at the same point, then $n_{\cos}$ defines the number of terms in this formula to get the value of $\tilde\Phi$ with the same accuracy. In the column for $n_{\cos}$, the dash means that the result of calculation is either incorrect (due to the computational difficulties) or indeterminate (because of the divergence of the function $K_0$). It is clear that the results of calculations depend on the ratio of the effective screening length $\lambda_{\mathrm{eff}}$ and the physical size $al$ of the period of the torus: $\tilde\lambda_{\mathrm{eff}}=\lambda_{\mathrm{eff}}/(al)$. Therefore, in Tables~\ref{results_table_1},~\ref{results_table_2} we present the numbers obtained for both small and large values of $\tilde\lambda_{\mathrm{eff}}$: 0.01, 0.1, 1 and 10, respectively.

\begin{table}[t]
	\centering
\begin{tabular}{|c|c|c|c|c|c|}
	\hline
	&\phantom{x}&\phantom{x}&\phantom{x}&\phantom{x}&\phantom{x}\\
	& $\tilde z$ & $\tilde \xi$ & $\tilde{\Phi}$ & $n_{\exp}$ & $n_{\cos}$ \\[5pt]
	
	\hline
	
	$A_1$ & 0.5 & 0.5 & $5.524\times10^{-31}$ & 2 & 20 \\
	
	\hline
	
	$A_2$ & 0.5 & 0.1 & $2.810\times10^{-22}$ & 2 & --- \\
	
	\hline
	
	$A_3$ & 0.5 & 0 & $7.715\times10^{-22}$ & 2 &--- \\
	
	\hline
	
	$B_1$ & 0.1 & 0.5 & $1.405\times10^{-22}$ & 1 & 10 \\
	
	\hline
	
	$B_2$ & 0.1 & 0.1 & $5.101\times10^{-6}$ & 1 & 31 \\
	
	\hline
	
	$B_3$ & 0.1 & 0 & $4.540\times10^{-4}$ & 1 &--- \\
	
	\hline
	
	$C_1$ & 0 & 0.5 & $3.857\times10^{-22}$ & 1 & 9 \\
	
	\hline
	
	$C_2$ & 0 & 0.1 & $4.540\times10^{-4}$ & 1 & 24 \\
	
	\hline
\end{tabular} \hspace{1cm}
\begin{tabular}{|c|c|c|c|c|c|}
	\hline
	&\phantom{x}&\phantom{x}&\phantom{x}&\phantom{x}&\phantom{x}\\
	& $\tilde z$ & $\tilde \xi$ & $\tilde{\Phi}$ & $n_{\exp}$ & $n_{\cos}$ \\[5pt]
	
	\hline
	
	$A_1$ & 0.5 & 0.5 & $2.402\times10^{-3}$ & 2 & 6 \\
	
	\hline
	
	$A_2$ & 0.5 & 0.1 & $2.394\times10^{-2}$ & 2 & 21\\
	
	\hline
	
	$A_3$ & 0.5 & 0 & $2.695\times10^{-2}$ & 2 &--- \\
	
	\hline
	
	$B_1$ & 0.1 & 0.5 & $1.201\times10^{-2}$ & 2 & 4 \\
	
	\hline
	
	$B_2$ & 0.1 & 0.1 & $1.719$ & 1 & 15 \\
	
	\hline
	
	$B_3$ & 0.1 & 0 & $3.679$ & 1 &--- \\
	
	\hline
	
	$C_1$ & 0 & 0.5 & $1.350\times10^{-2}$ & 2 & 4 \\
	
	\hline
	
	$C_2$ & 0 & 0.1 & $3.679$ & 1 & 15 \\
	
	\hline
\end{tabular}

\vspace{0.3cm}

\caption{\label{results_table_1}Values of the rescaled gravitational potential $\tilde\Phi$ and corresponding numbers $n_{\exp}$ and $n_{\cos}$  of terms of series for some selected points in the cases $\tilde\lambda_{\mathrm{eff}}=0.01$ and $\tilde\lambda_{\mathrm{eff}}=0.1$ for the left and right tables, respectively.}
\end{table}

\begin{table}[t]
	\centering
\begin{tabular}{|c|c|c|c|c|c|}
	\hline
	&\phantom{x}&\phantom{x}&\phantom{x}&\phantom{x}&\phantom{x}\\	
	& $\tilde z$ & $\tilde \xi$ & $\tilde{\Phi}$ & $n_{\exp}$ & $n_{\cos}$ \\[5pt]
	
	\hline
	
	$A_1$ & 0.5 & 0.5 & $1.740$ & 11 & 3 \\
	
	\hline
	
	$A_2$ & 0.5 & 0.1 & $2.741$ & 7 & 14\\
	
	\hline
	
	$A_3$ & 0.5 & 0 & $2.814$ & 8 &--- \\
	
	\hline
	
	$B_1$ & 0.1 & 0.5 & $1.941$ & 7 & 3 \\
	
	\hline
	
	$B_2$ & 0.1 & 0.1 & $7.068$ & 7 & 15 \\
	
	\hline
	
	$B_3$ & 0.1 & 0 & $9.986$ & 7 &--- \\
	
	\hline
	
	$C_1$ & 0 & 0.5 & $1.965$ & 7 & 3 \\
	
	\hline
	
	$C_2$ & 0 & 0.1 & $9.958$ & 8 & 15 \\
	
	\hline
\end{tabular}\hspace{1cm}
\begin{tabular}{|c|c|c|c|c|c|}
	\hline
	&\phantom{x}&\phantom{x}&\phantom{x}&\phantom{x}&\phantom{x}\\
	& $\tilde z$ & $\tilde \xi$ & $\tilde{\Phi}$ & $n_{\exp}$ & $n_{\cos}$ \\[5pt]
	
	\hline
	
	$A_1$ & 0.5 & 0.5 & $6.114$ & 64 & 3 \\
	
	\hline
	
	$A_2$ & 0.5 & 0.1 & $7.297$ & 81 & 15\\
	
	\hline
	
	$A_3$ & 0.5 & 0 & $7.378$ & 62 &--- \\
	
	\hline
	
	$B_1$ & 0.1 & 0.5 & $6.325$ & 64 & 3 \\
	
	\hline
	
	$B_2$ & 0.1 & 0.1 & $11.69$ & 49 & 11 \\
	
	\hline
	
	$B_3$ & 0.1 & 0 & $14.63$ & 46 & ---\\
	
	\hline
	
	$C_1$ & 0 & 0.5 & $6.350$ & 61 & 3 \\
	
	\hline
	
	$C_2$ & 0 & 0.1 & $14.59$ & 40 & 10 \\
	
	\hline
\end{tabular}

\vspace{0.3cm}

\caption{\label{results_table_2}Values of the rescaled gravitational potential $\tilde\Phi$ and corresponding numbers $n_{\exp}$ and $n_{\cos}$  of terms of series for some selected points in the cases $\tilde\lambda_{\mathrm{eff}}=1$ and $\tilde\lambda_{\mathrm{eff}}=10$ for the left and right tables, respectively.}
\end{table}

These tables demonstrate that the formula \rf{3.2} with Yukawa potentials generally requires much less number of terms ($n_{\exp} \ll n_{\cos}$) when the screening length is less than the period of the torus, {\it i.e.} $\tilde\lambda_{\mathrm{eff}} < 1$. As we mentioned in Introduction, the lower limit on the period of the torus following from the observations is of the order of 16~Gpc~\cite{42}. On the other hand, the effective cosmological screening length at the present time is 2.6~Gpc \cite{MaxEzgi}. Therefore, the inequality  $\tilde\lambda_{\mathrm{eff}} < 1$ corresponds to the observable Universe, and here eq.~\rf{3.2} is preferable for the numerical analysis.

\begin{figure*}
	\resizebox{0.48\textwidth}{!}{\includegraphics{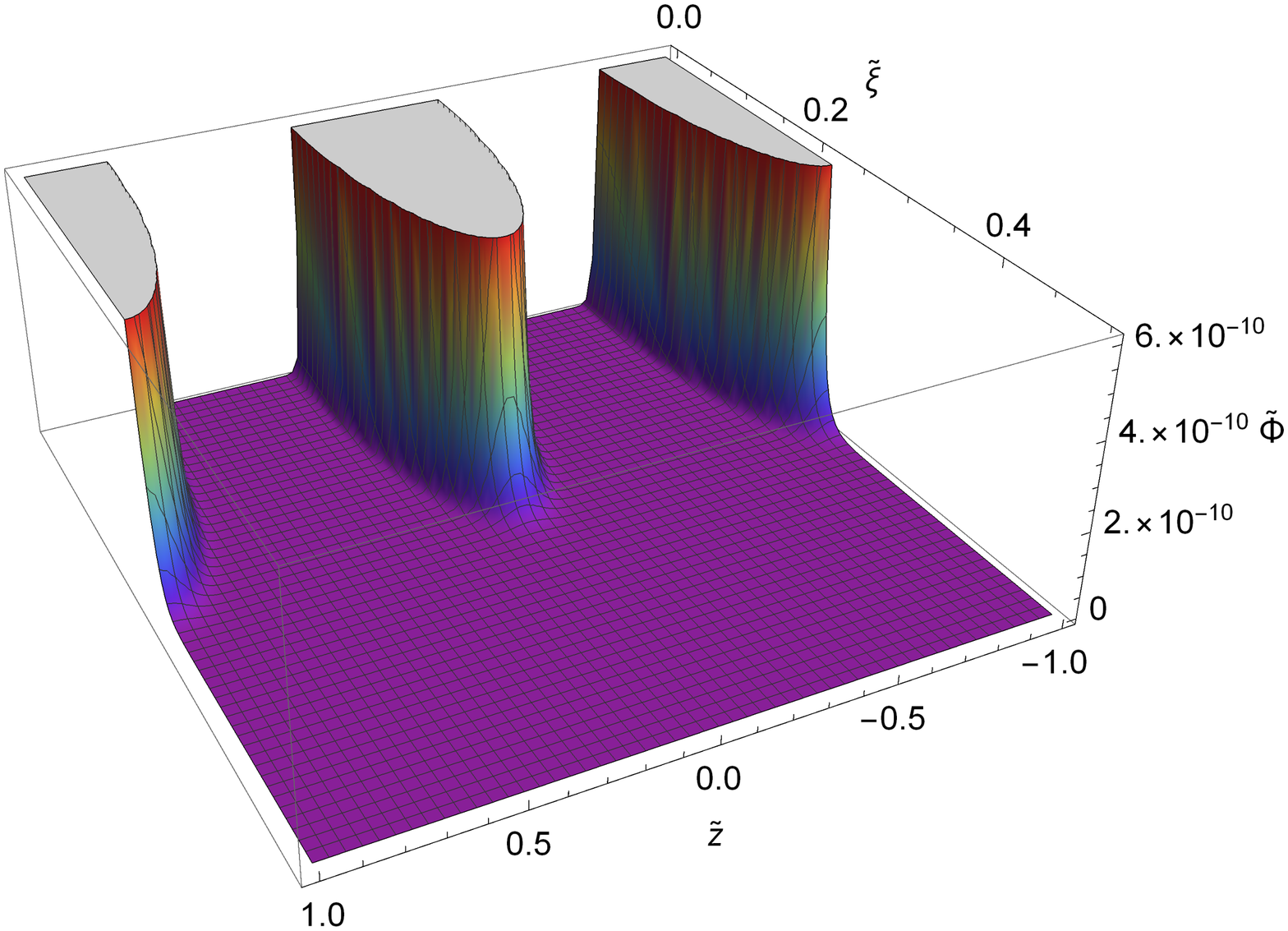}}\quad\quad
	\resizebox{0.48\textwidth}{!}{\includegraphics{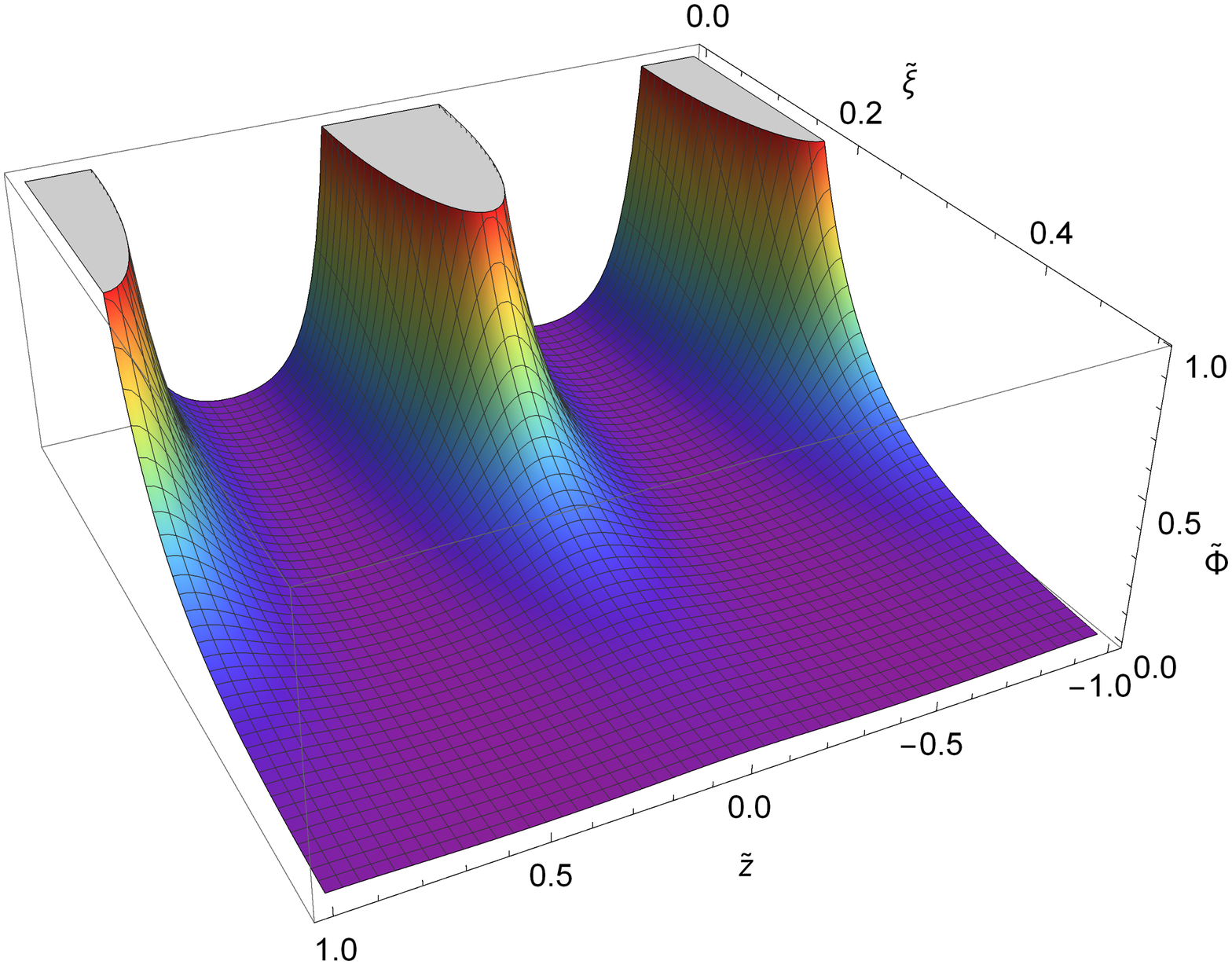}}
	\caption{Rescaled gravitational potential  $\tilde\Phi=\left[-G_N m/(c^2al)\right]^{-1}\widehat\Phi$ in the cases $\tilde\lambda_{\mathrm{eff}} = 0.01$ and $\tilde\lambda_{\mathrm{eff}} = 0.1$ for the left and right panels, respectively.}   
	\label{fig:1}
%\end{figure*}
\vspace{0.6cm}
%\begin{figure*}
	\resizebox{0.48\textwidth}{!}{\includegraphics{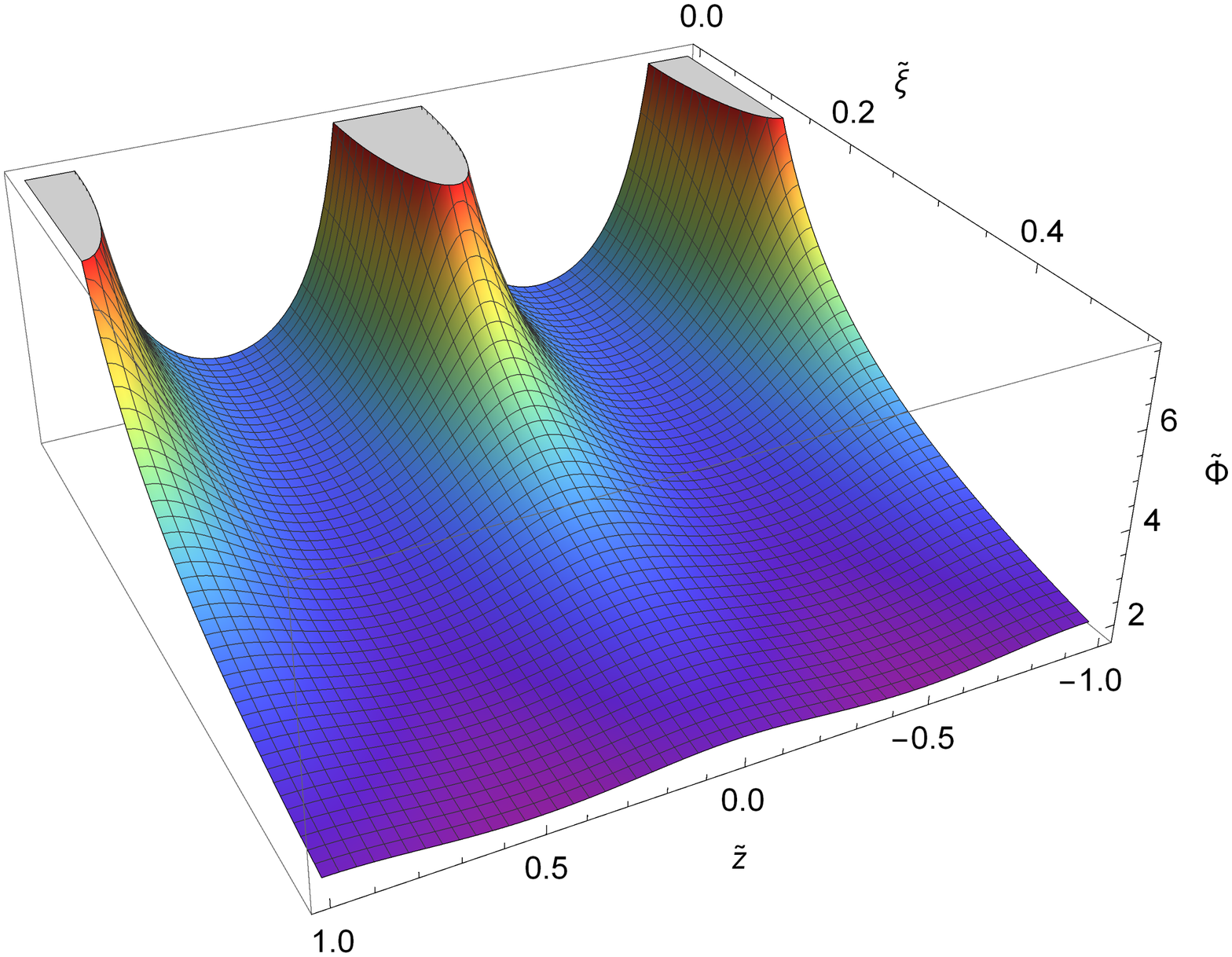}}\quad\quad
	\resizebox{0.48\textwidth}{!}{\includegraphics{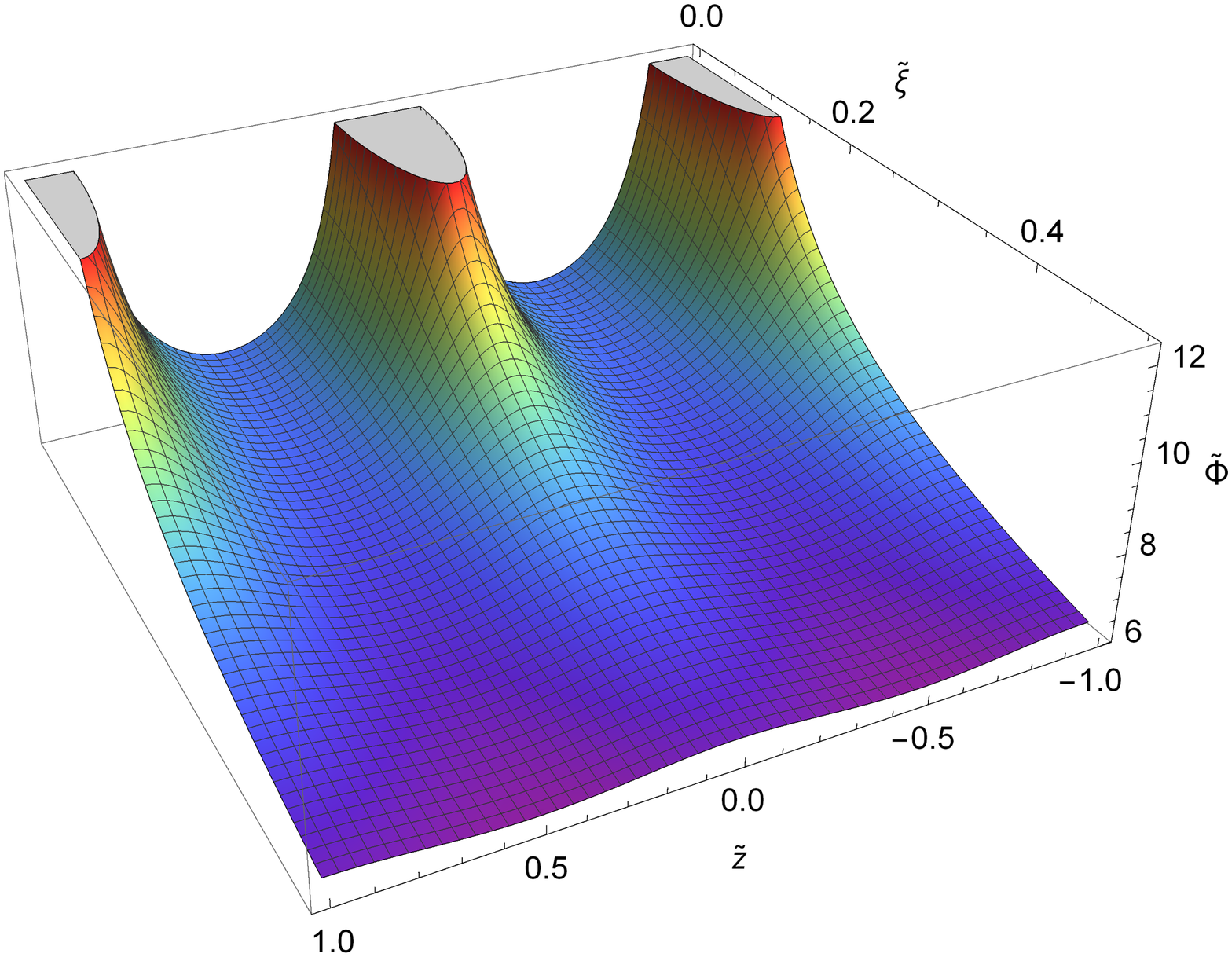}}
	\caption{Rescaled gravitational potential  $\tilde\Phi=\left[-G_N m/(c^2al)\right]^{-1}\widehat\Phi$ in the cases $\tilde\lambda_{\mathrm{eff}} = 1$ and $\tilde\lambda_{\mathrm{eff}} = 10$ for the left and right panels, respectively.}   
	\label{fig:2}
\end{figure*}

To conclude this section, we present Figs.~\ref{fig:1},~\ref{fig:2} of the rescaled gravitational potential $\tilde\Phi$ which correspond to four different values of $\tilde\lambda_{\mathrm{eff}}$ selected for Tables~\ref{results_table_1},~\ref{results_table_2}. To draw these pictures (with the help of Mathematica \cite{Math}), we use the formula \rf{3.2} where we  choose  $n \gg n_{\exp}$.
%For example, $n = 10, 100, 1000, 2000$ for 
%$\tilde\lambda_{\mathrm{eff}}= 0.01, 0.1, 1, 10$, respectively.

%\vspace{5cm}

%%%%%%%%%%%%%%\prod
%\begin{figure*}[htbp]
%	\includegraphics[width=0.5in,height=0.5in]{gr2.pdf}
%		\%includegraphics[width=0.5in,height=0.5in]{gr3.pdf}
%	\caption {Dynamics of three gravitating masses . \%label{triangle}}
%\end{figure*}
%%%%%%%%%%%%%%%%%%%%%%%%%%%%

%%%%%%%%%%%%%%%%%%%%%%%%%%%%%%%%%%%%%%%%
%%%%%%%%%%%%%%%%%%%%%%%%%%%%%%%%%%%%%%%%%%%
%%%%%%%%%%%%%%%%%%%%%%%%%%%%%%%%%%%%%%%%%%

\section{Gravitational forces}\label{sec:4}

\setcounter{equation}{0}

Below we present the gravitational forces (per unit mass)
corresponding to the potentials considered in the previous section. More precisely, we calculate the projections of these forces on the $z$-axis and on the polar radius $\xi$ for the selected points. Among these points, it is natural to consider only those ones where the corresponding projections are nonzero. Similarly to the potentials, we calculate these projections up to the fourth digit and find numbers of terms of the series starting from which we achieve this precision. The less the number of terms, the better the formula for the numerical analysis. Thus, we compare two alternative expressions following from \rf{2.29} and \rf{2.30} from this point of view.

\subsection{$z$-component of the gravitational force}\label{sec:4.1}
%%%%%%%%%%%%%%%%%%%%%%%%%%%%%%%%%%%%%%%%%
First, we consider $z$-components of the gravitational forces. From the potentials \rf{2.29} and \rf{2.30}
we obtain two alternative formulas for the rescaled $z$-component of the force:
%%%%%%%
\be{4.1}
\left.\frac{\partial}{\partial\tilde z}\left(\tilde{\Phi}_{\cos}\right)\right|_{n_{\cos}}=
-8\pi\sum_{k=1}^{n_{\cos}}
kK_0\left(\sqrt{4\pi^2k^2+\frac{1}{\tilde\lambda_{\mathrm{eff}}^2}}\tilde\xi\right) \sin(2\pi k \tilde{z})\, 
\ee
%%%%%%%%
and
%%%%%%%
\ba{4.2} &{}&\left.\frac{\partial}{\partial\tilde z}\left(\tilde{\Phi}_{\exp}\right)\right|_{n_{\exp}}=
-\left(\frac{\tilde{z}}{\tilde\lambda_{\mathrm{eff}}\left(\tilde\xi^2+\tilde z^2\right)}
+ \frac{\tilde{z}}{\left(\tilde\xi^2+\tilde{z}^2\right)^{3/2}}\right)\exp\left(-\frac{\sqrt{\tilde\xi^2 + \tilde{z}^2}}{\lambda_{\mathrm{eff}}}\right)\nn\\ 
&-&  \sum_{k=1}^{n_{\exp}-1}
\left[\left(\frac{\tilde{z}+k}{\tilde\lambda_{\mathrm{eff}}\left(\tilde\xi^2+(\tilde{z}+k)^2\right)}
+ \frac{\tilde{z}+k}{\left(\tilde\xi^2+(\tilde{z}+k)^2\right)^{3/2}}\right) \exp\left(-\frac{\sqrt{\tilde\xi^2+(\tilde{z}+k)^2}}{\tilde\lambda_{\mathrm{eff}}}\right)\right.\nn\\ 
&+&\left.
\left(\frac{\tilde{z}-k}{\tilde\lambda_{\mathrm{eff}}\left(\tilde\xi^2+(\tilde{z}-k)^2\right)}
+ \frac{\tilde{z}-k}{\left(\tilde\xi^2+(\tilde{z}-k)^2\right)^{3/2}}\right)\exp\left(-\frac{\sqrt{\tilde\xi^2+(\tilde{z}-k)^2}}{\tilde\lambda_{\mathrm{eff}}}\right)
\right]\, .\ea
%%%%%%%%

Obviously, these $z$-components are equal to zero at the points $A_1, A_2, A_3$, $C_1$ and $C_2$. Therefore, we consider only the points $B_1, B_2$ and $B_3$. The results of our computations with the help of Mathematica \cite{Math} are presented in Tables~\ref{results_table_3},~\ref{results_table_4} for four values of $\tilde\lambda_{\mathrm{eff}}$: 0.01, 0.1, 1 and 10, respectively.  These results demonstrate that, similarly to the gravitational potential, the formula \rf{4.2}, based on the Yukawa potentials, is preferable for the physically relevant case $\tilde\lambda_{\mathrm{eff}} < 1$. In these tables, the values of the rescaled $z$-component $\tilde{\Phi}_z$ are computed with the help of eq.~\rf{4.2} for $n\gg n_{\exp}$.
% $\tilde{\Phi}_z\equiv \left.\frac{\partial}{\partial\tilde
% z}\left(\tilde{\Phi}_{\exp}\right)\right|_{n>>n_{\exp}}$.

%%%%%%%%%%%%%%%%%%%%%%%%%%%%%%%%%%%%%%%%%%%%%%%%%%%%%%
%%%%%%%%%%%%%%%%%%%%%%%%%%%%%%%%%%%%%%%%%%%%%%%%%%%%%%
\begin{table}[t]
	\centering
	\begin{tabular}{|c|c|c|c|c|c|}
		\hline
		&\phantom{x}&\phantom{x}&\phantom{x}&\phantom{x}&\phantom{x}\\
	& $\tilde z$ & $\tilde \xi$ & $\tilde{\Phi}_z$ & $n_{\exp}$ & $n_{\cos}$  \\[5pt]
	
	\hline

	$B_1$ & 0.1 & 0.5 & $-2.810\times10^{-21}$ & 1 & 9 \\
	
	\hline
	
	$B_2$ & 0.1 & 0.1 & $-3.862\times10^{-4}$ & 1 & 32 \\
	
	\hline
	
	$B_3$ & 0.1 & 0 & $-4.994\times10^{-2}$ & 1 & ---  \\
	
	\hline

\end{tabular} \hspace{1cm}
\begin{tabular}{|c|c|c|c|c|c|}
	\hline
	&\phantom{x}&\phantom{x}&\phantom{x}&\phantom{x}&\phantom{x}\\
	& $\tilde z$ & $\tilde \xi$ & $\tilde{\Phi}_z$ & $n_{\exp}$ & $n_{\cos}$  \\[5pt]
	
	\hline

	$B_1$ & 0.1 & 0.5 & $-2.781\times10^{-2}$ & 2 & 4 \\
	
	\hline
	
	$B_2$ & 0.1 & 0.1 & $-20.75$ & 1 & 17\\
	
	\hline
	
	$B_3$ & 0.1 & 0 & $-73.57$ & 2 & --- \\
	
	\hline

\end{tabular}

\vspace{0.3cm}

\caption{\label{results_table_3}Values of the rescaled $z$-component of the gravitational force $\tilde{\Phi}_z$ and corresponding numbers $n_{\exp}$ and $n_{\cos}$  of terms of series for points $B_1, B_2$ and $B_3$ in the cases $\tilde\lambda_{\mathrm{eff}}=0.01$ and $\tilde\lambda_{\mathrm{eff}}=0.1$ for the left and right tables, respectively.}
\end{table}
%%%%%%%%%%%%%%%%%%%%%%%%%%%%%%%%%%%%%%%%%%%%%
%%%%%%%%%%%%%%%%%%%%%%%%%%%%%%%%%%%%%%%%%%%%

\begin{table}[t]
	\centering
	\begin{tabular}{|c|c|c|c|c|c|}
		\hline
		&\phantom{x}&\phantom{x}&\phantom{x}&\phantom{x}&\phantom{x}\\
		& $\tilde z$ & $\tilde \xi$ & $\tilde{\Phi}_z$ & $n_{\exp}$ & $n_{\cos}$  \\[5pt]
	
	\hline

	$B_1$ & 0.1 & 0.5 & $-4.618\times10^{-1}$ & 8 & 3  \\
	
	\hline
	
	$B_2$ & 0.1 & 0.1 & $-34.63$ & 4 & 16  \\
	
	\hline
	
	$B_3$ & 0.1 & 0 & $-99.11$ & 5 & ---  \\
	
	\hline

\end{tabular} \hspace{1cm}
\begin{tabular}{|c|c|c|c|c|c|}
\hline
&\phantom{x}&\phantom{x}&\phantom{x}&\phantom{x}&\phantom{x}\\
& $\tilde z$ & $\tilde \xi$ & $\tilde{\Phi}_z$ & $n_{\exp}$ & $n_{\cos}$  \\[5pt]

	\hline

	$B_1$ & 0.1 & 0.5 & $-4.819\times10^{-1}$ & 39 & 3\\
	
	\hline
	
	$B_2$ & 0.1 & 0.1 & $-34.88$ & 6 & 16  \\
	
	\hline
	
	$B_3$ & 0.1 & 0 & $-99.51$ & 6 & ---  \\
	
	\hline

\end{tabular}

\vspace{0.3cm}

\caption{\label{results_table_4}Values of the rescaled $z$-component of the gravitational force $\tilde{\Phi}_z$ and corresponding numbers $n_{\exp}$ and $n_{\cos}$  of terms of series for points $B_1, B_2$ and $B_3$ in the cases $\tilde\lambda_{\mathrm{eff}}=1$ and $\tilde\lambda_{\mathrm{eff}}=10$ for the left and right tables, respectively.}
\end{table}
%%%%%%%%%%%%%%%%%%%%%%%%%%%%%%%%%%%%%%%%%%%%%
%%%%%%%%%%%%%%%%%%%%%%%%%%%%%%%%%%%%%%%%%%%%

\begin{figure*}
	\resizebox{0.48\textwidth}{!}{\includegraphics{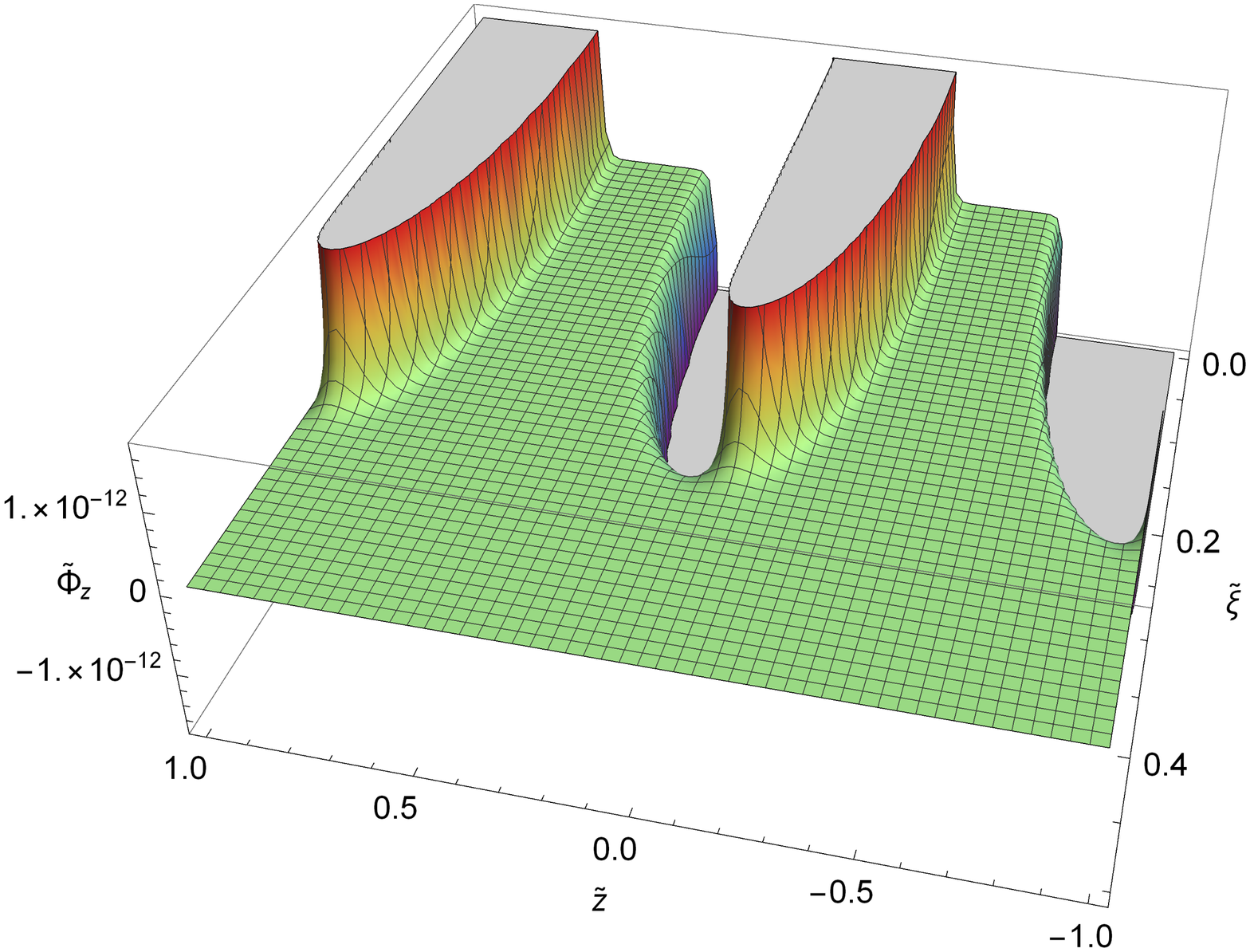}}\quad\quad
	\resizebox{0.48\textwidth}{!}{\includegraphics{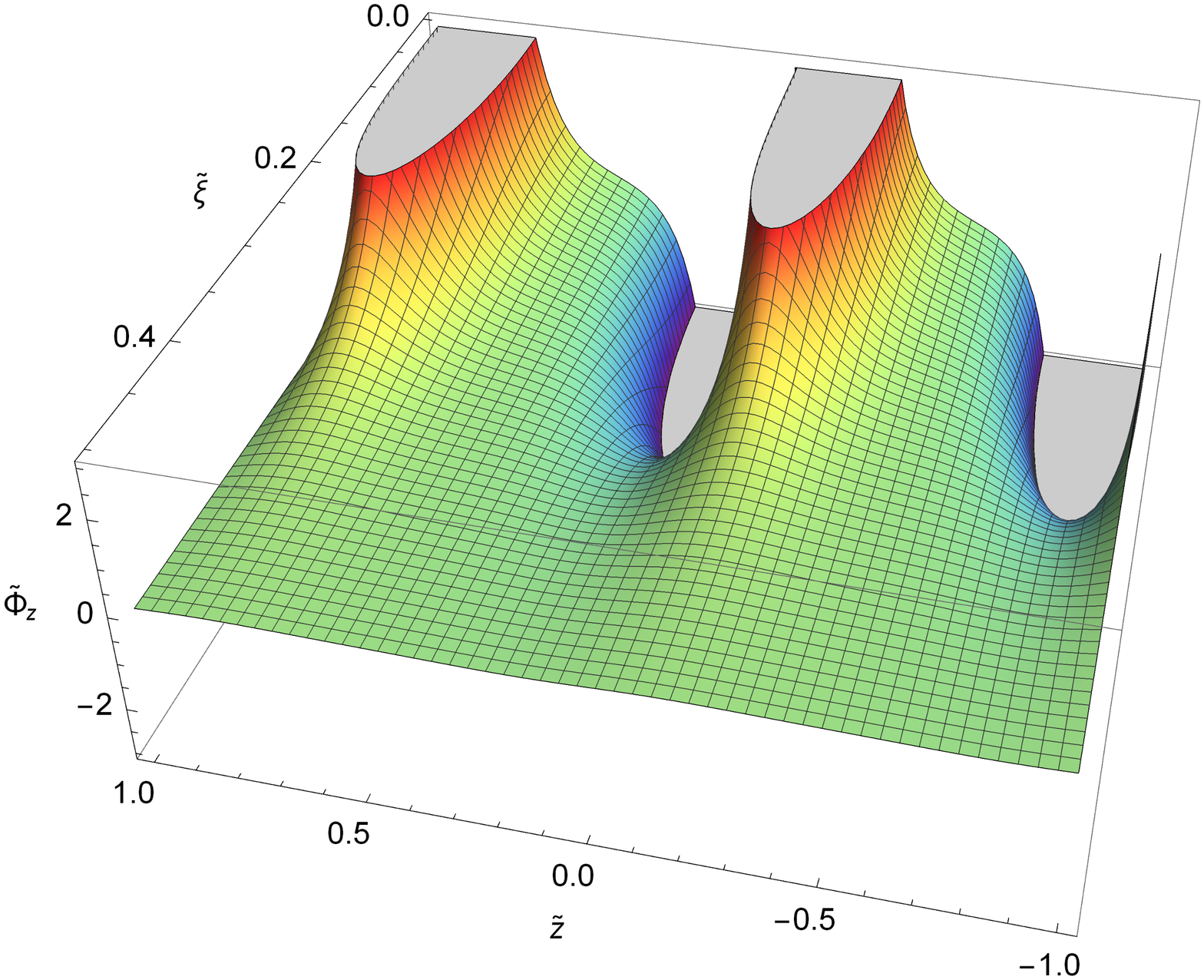}}
	\caption{Rescaled $z$-component of the gravitational force $\tilde{\Phi}_z\equiv \partial \tilde\Phi/\partial z$ in the cases $\tilde\lambda_{\mathrm{eff}} = 0.01$ and $\tilde\lambda_{\mathrm{eff}} = 0.1$ (left and right panels, respectively).}   
	\label{fig:3}
%\end{figure*}
\vspace{0.6cm}
%\begin{figure*}
	\resizebox{0.48\textwidth}{!}{\includegraphics{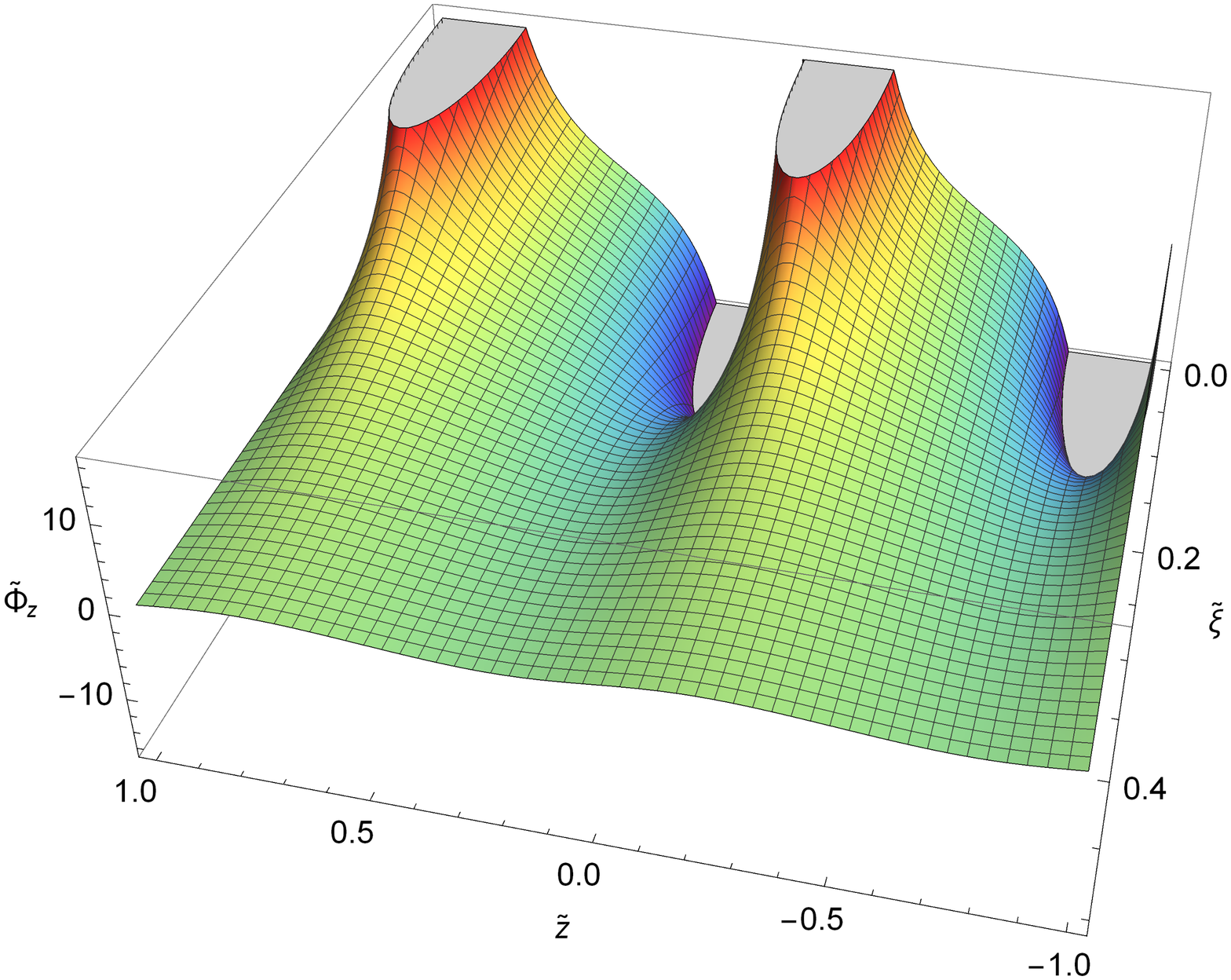}}\quad\quad
	\resizebox{0.48\textwidth}{!}{\includegraphics{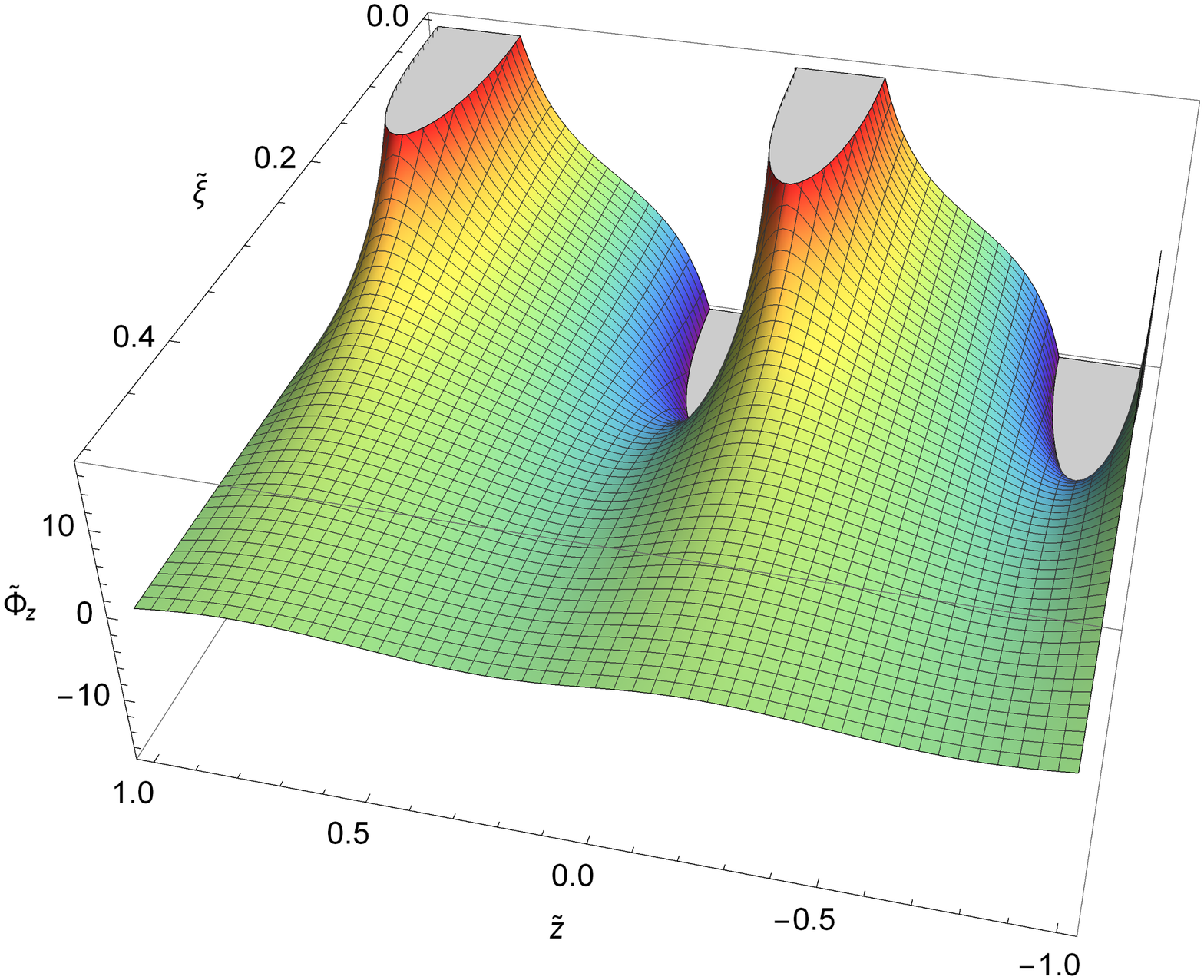}}
	\caption{Rescaled $z$-component of the gravitational force $\tilde{\Phi}_z\equiv \partial \tilde\Phi/\partial z$ in the cases $\tilde\lambda_{\mathrm{eff}} = 1$ and $\tilde\lambda_{\mathrm{eff}} = 10$ (left and right panels, respectively).}   
	\label{fig:4}
\end{figure*}

In addition, we present Figs.~\ref{fig:3},~\ref{fig:4} of the rescaled $z$-components of the gravitational force $\tilde{\Phi}_z$ which correspond to four different values of $\tilde\lambda_{\mathrm{eff}}$ selected for Tables~\ref{results_table_3},~\ref{results_table_4}. To draw these pictures (with the help of Mathematica \cite{Math}), we use the formula \rf{4.2} where we  choose  $n \gg n_{\exp}$. 
%For example, $n = 10, 100, 1000, 2000$ for   
%$\tilde\lambda_{\mathrm{eff}}= 0.01, 0.1, 1, 10$, %respectively.

%%%%%%%%%%%%%%%%%%%%%%%%%%%%

%%%%%%%%%%%%%%%%%%%%%%%%%%%%%%%%%%%%%%%%
%%%%%%%%%%%%%%%%%%%%%%%%%%%%%%%%%%%%%%%%%%%
%%%%%%%%%%%%%%%%%%%%%%%%%%%%%%%%%%%%%%%%%%

\subsection{$\xi$-component of the gravitational force}\label{sec:4.2}
%%%%%%%%%%%%%%%%%%%%%%%%%%%%%%%%%%%%%%%%%

Now we turn to the $\xi$-component of the gravitational force. In this case, two alternative formulas are:
%%%%%%
\ba{4.3} \left.\frac{\partial}{\partial\tilde \xi}\left(\tilde{\Phi}_{\cos}\right)\right|_{n_{\cos}}=-\frac{2}{\tilde\lambda_{\mathrm{eff}}}K_1\left(\frac{\tilde\xi}{\tilde\lambda_{\mathrm{eff}}}\right)-4\sum_{k=1}^{n_{\cos}-1}\sqrt{4\pi^2k^2+\frac{1}{\tilde\lambda^2_{\mathrm{eff}}}}
{K}_{1}\left(\sqrt{4\pi^2k^2+\frac{1}{\tilde\lambda^2_{\mathrm{eff}}}}\tilde\xi\right)\cos\left(2\pi k\tilde{z}\right)\, \ea
%%%%%%%
and
%%%%%%%%
\ba{4.4}  &{}&\left.\frac{\partial}{\partial\tilde \xi}\left(\tilde{\Phi}_{\exp}\right)\right|_{n_{\exp}} =
-\left(\frac{\tilde\xi}{\tilde\lambda_{\mathrm{eff}}\left(\tilde\xi^2+\tilde z^2\right)}
+ \frac{\tilde\xi}{\left(\tilde\xi^2+\tilde{z}^2\right)^{3/2}}\right)\exp\left(-\frac{\sqrt{\tilde\xi^2 + \tilde{z}^2}}{\lambda_{\mathrm{eff}}}\right)\nn\\ 
&-&  \sum_{k=1}^{n_{\exp}-1}
\left[\left(\frac{\tilde\xi}{\tilde\lambda_{\mathrm{eff}}\left(\tilde\xi^2+(\tilde{z}+k)^2\right)}
+ \frac{\tilde\xi}{\left(\tilde\xi^2+(\tilde{z}+k)^2\right)^{3/2}}\right) \exp\left(-\frac{\sqrt{\tilde\xi^2+(\tilde{z}+k)^2}}{\tilde\lambda_{\mathrm{eff}}}\right)\right.\nn\\ 
&+&\left.
\left(\frac{\tilde\xi}{\tilde\lambda_{\mathrm{eff}}\left(\tilde\xi^2+(\tilde{z}-k)^2\right)}
+ \frac{\tilde\xi}{\left(\tilde\xi^2+(\tilde{z}-k)^2\right)^{3/2}}\right)\exp\left(-\frac{\sqrt{\tilde\xi^2+(\tilde{z}-k)^2}}{\tilde\lambda_{\mathrm{eff}}}\right)
\right]\, .\ea
%%%%%%%

Evidently, this component is equal to zero at the points with $\tilde\xi=0$ and nonzero $\tilde z$, for instance, at the points $A_3$ and $B_3$. Therefore, we consider only the points $A_1, A_2, B_1, B_2, C_1$ and $C_2$. The results of numerical calculations with the help of Mathematica \cite{Math}, based on the formulas \rf{4.3} and \rf{4.4}, for four values of $\tilde\lambda_{\mathrm{eff}}$ are presented in Tables~\ref{results_table_5},~\ref{results_table_6}. We calculate the $\xi$-component up to the fourth digit and find how many terms $n_{\cos}$ and $n_{\exp}$ of the series in eqs.~\rf{4.3} and \rf{4.4} are required for this purpose.  $\tilde{\Phi}_{\xi}$ denotes the values of the rescaled $\xi$-component computed with the help of eq.~\rf{4.4} for $n\gg n_{\exp}$.
%: $\; \tilde{\Phi}_{\xi}\equiv \left.\frac{\partial}{\partial\tilde \xi}
%\left(\tilde{\Phi}_{\exp}\right)\right|_{n>>n_{\exp}}$. 
These tables, similarly to Tables~\ref{results_table_1}-\ref{results_table_4}, demonstrate that in the case $\tilde\lambda_{\mathrm{eff}}<1$, which corresponds to the observational restrictions, the formula \rf{4.4} is preferable.

%%%%%%%%%%%%%%%%%%%%%%%%%%%%%%%%%%%%%%%%%%%%%
%%%%%%%%%%%%%%%%%%%%%%%%%%%%%%%%%%%%%%%%%%%%

\begin{table}[t]
	\centering
	\begin{tabular}{|c|c|c|c|c|c|}
		\hline
		&\phantom{x}&\phantom{x}&\phantom{x}&\phantom{x}&\phantom{x}\\
		& $\tilde z$ & $\tilde \xi$ & $\tilde{\Phi}_{\xi}$ & $n_{\exp}$ & $n_{\cos}$  \\[5pt]
	
	\hline
	
	$A_1$ & 0.5 & 0.5 & $-3.962\times10^{-29}$ & 2 & 20\\
	
	\hline
	
	$A_2$ & 0.5 & 0.1 &  $-5.620\times10^{-21}$ & 2 & ---\\
	
	\hline

	$B_1$ & 0.1 & 0.5 &  $-1.405\times10^{-20}$ & 1 & 10\\
	
	\hline
	
	$B_2$ & 0.1 & 0.1 & $-3.862\times10^{-4}$ & 1 & 32\\
	
	\hline

	$C_1$ & 0 & 0.5 &  $-3.935\times10^{-20}$ & 1 & 11 \\
	
	\hline
	
	$C_2$ & 0 & 0.1 &  $-4.994\times10^{-2}$ & 1 & 25 \\
	
	\hline
	
\end{tabular} \hspace{1cm}
\begin{tabular}{|c|c|c|c|c|c|}
\hline
&\phantom{x}&\phantom{x}&\phantom{x}&\phantom{x}&\phantom{x}\\
& $\tilde z$ & $\tilde \xi$ & $\tilde{\Phi}_{\xi}$ & $n_{\exp}$ & $n_{\cos}$  \\[5pt]
	
	\hline
	
	$A_1$ & 0.5 & 0.5 &  $-1.939\times10^{-2}$ & 2 & 5\\
	
	\hline
	
	$A_2$ & 0.5 & 0.1 &  $-5.615\times10^{-2}$ & 2 & 28\\
	
	\hline

	$B_1$ & 0.1 & 0.5 &  $-1.406\times10^{-1}$ & 2 & 4\\
	
	\hline
	
	$B_2$ & 0.1 & 0.1 &  $-20.75$ & 1 & 17\\
	
	\hline

	$C_1$ & 0 & 0.5 &  $-1.618\times10^{-1}$ & 2 & 4\\
	
	\hline
	
	$C_2$ & 0 & 0.1 &  $-73.58$ & 1 & 21\\
	
	\hline
	
\end{tabular}

\vspace{0.3cm}

\caption{\label{results_table_5}Values of the rescaled $\xi$-component of the gravitational force $\tilde{\Phi}_{\xi}$ and corresponding numbers $n_{\exp}$ and $n_{\cos}$  of terms of series for points $A_1, A_2, B_1, B_2, C_1$ and $C_2$ in the cases $\tilde\lambda_{\mathrm{eff}}=0.01$ and $\tilde\lambda_{\mathrm{eff}}=0.1$ for the left and right tables, respectively.}
\end{table}
%%%%%%%%%%%%%%%%%%%%%%%%%%%%%%%%%%%%%%%%%%%%%
%%%%%%%%%%%%%%%%%%%%%%%%%%%%%%%%%%%%%%%%%%%%

\begin{table}[t]
	\centering
	\begin{tabular}{|c|c|c|c|c|c|}
		\hline
		&\phantom{x}&\phantom{x}&\phantom{x}&\phantom{x}&\phantom{x}\\
		& $\tilde z$ & $\tilde \xi$ & $\tilde{\Phi}_{\xi}$ & $n_{\exp}$ & $n_{\cos}$  \\[5pt]
	
	\hline
	
	$A_1$ & 0.5 & 0.5 & $-2.536$ & 6 & 4 \\
	
	\hline
	
	$A_2$ & 0.5 & 0.1 &  $-1.405$ & 4 & 20 \\
	
	\hline

	$B_1$ & 0.1 & 0.5 &  $-3.993$ & 5 & 4 \\
	
	\hline
	
	$B_2$ & 0.1 & 0.1 &  $-35.20$ & 3 & 17 \\
	
	\hline

	$C_1$ & 0 & 0.5 &  $-4.188$ & 6 & 4 \\
	
	\hline
	
	$C_2$ & 0 & 0.1 &  $-99.69$ & 3 & 19\\
	
	\hline
	
\end{tabular} \hspace{1cm}
\begin{tabular}{|c|c|c|c|c|c|}
\hline
&\phantom{x}&\phantom{x}&\phantom{x}&\phantom{x}&\phantom{x}\\
& $\tilde z$ & $\tilde \xi$ & $\tilde{\Phi}_{\xi}$ & $n_{\exp}$ & $n_{\cos}$  \\[5pt]

	\hline
	
	$A_1$ & 0.5 & 0.5 &  $-3.177$ & 17 & 4 \\
	
	\hline
	
	$A_2$ & 0.5 & 0.1 &  $-1.587$ & 12 & 21 \\
	
	\hline

	$B_1$ & 0.1 & 0.5 &  $-4.686$ & 23 & 4 \\
	
	\hline
	
	$B_2$ & 0.1 & 0.1 &  $-35.60$ & 6 & 16 \\
	
	\hline

	$C_1$ & 0 & 0.5 &  $-4.886$ & 16 & 4 \\
	
	\hline
	
	$C_2$ & 0 & 0.1 &  $-100.2$ & 2 & 14\\
	
	\hline
	
\end{tabular}

\vspace{0.3cm}

\caption{\label{results_table_6}Values of the rescaled $\xi$-component of the gravitational force $\tilde{\Phi}_{\xi}$ and corresponding numbers $n_{\exp}$ and $n_{\cos}$  of terms of series for points $A_1, A_2, B_1, B_2, C_1$ and $C_2$ in the cases $\tilde\lambda_{\mathrm{eff}}=1$ and $\tilde\lambda_{\mathrm{eff}}=10$ for the left and right tables, respectively.}
\end{table}
%%%%%%%%%%%%%%%%%%%%%%%%%%%%%%%%%%%%%%%%%%%%%
%%%%%%%%%%%%%%%%%%%%%%%%%%%%%%%%%%%%%%%%%%%%

\begin{figure*}
	\resizebox{0.48\textwidth}{!}{\includegraphics{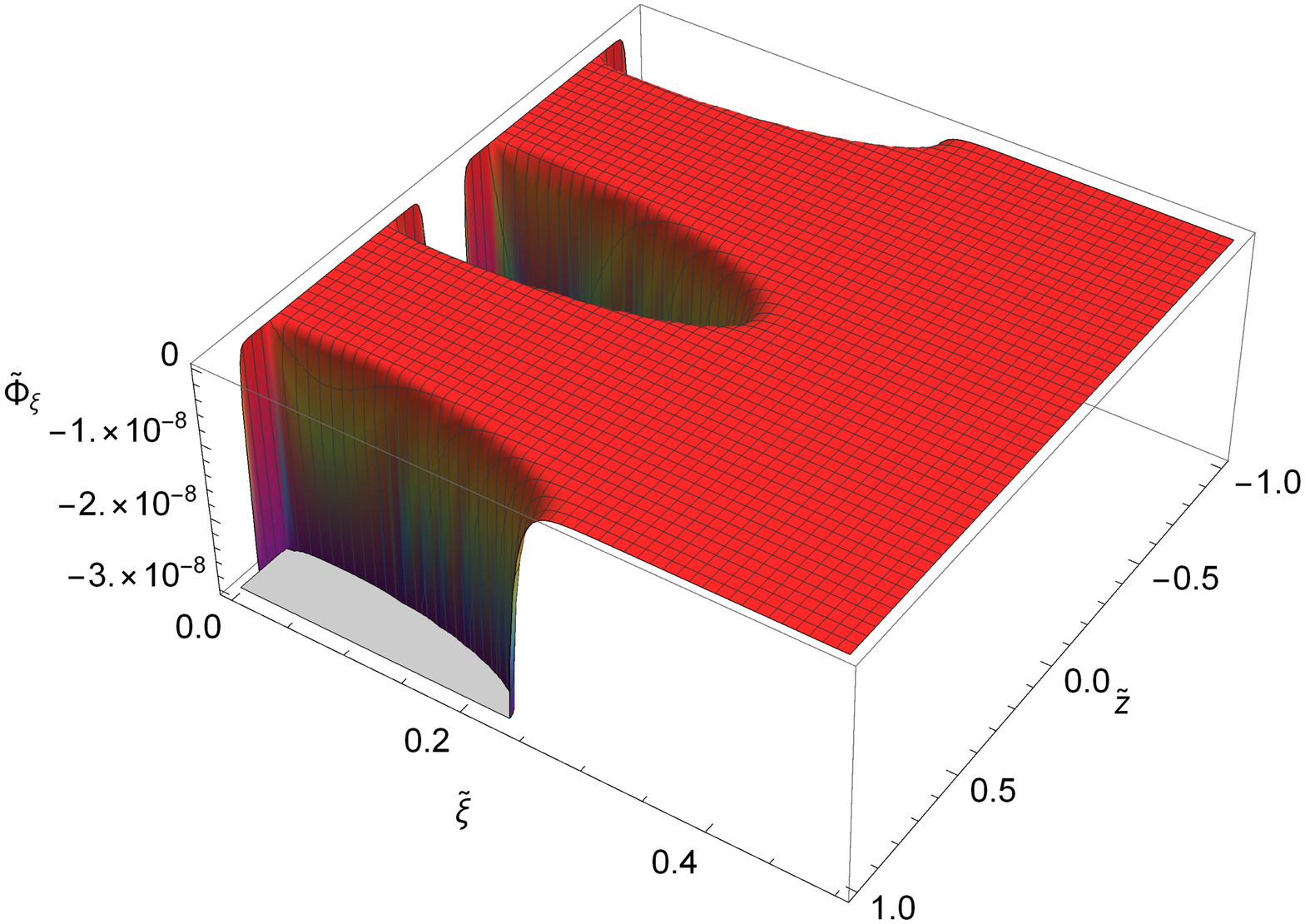}}\quad\quad
	\resizebox{0.48\textwidth}{!}{\includegraphics{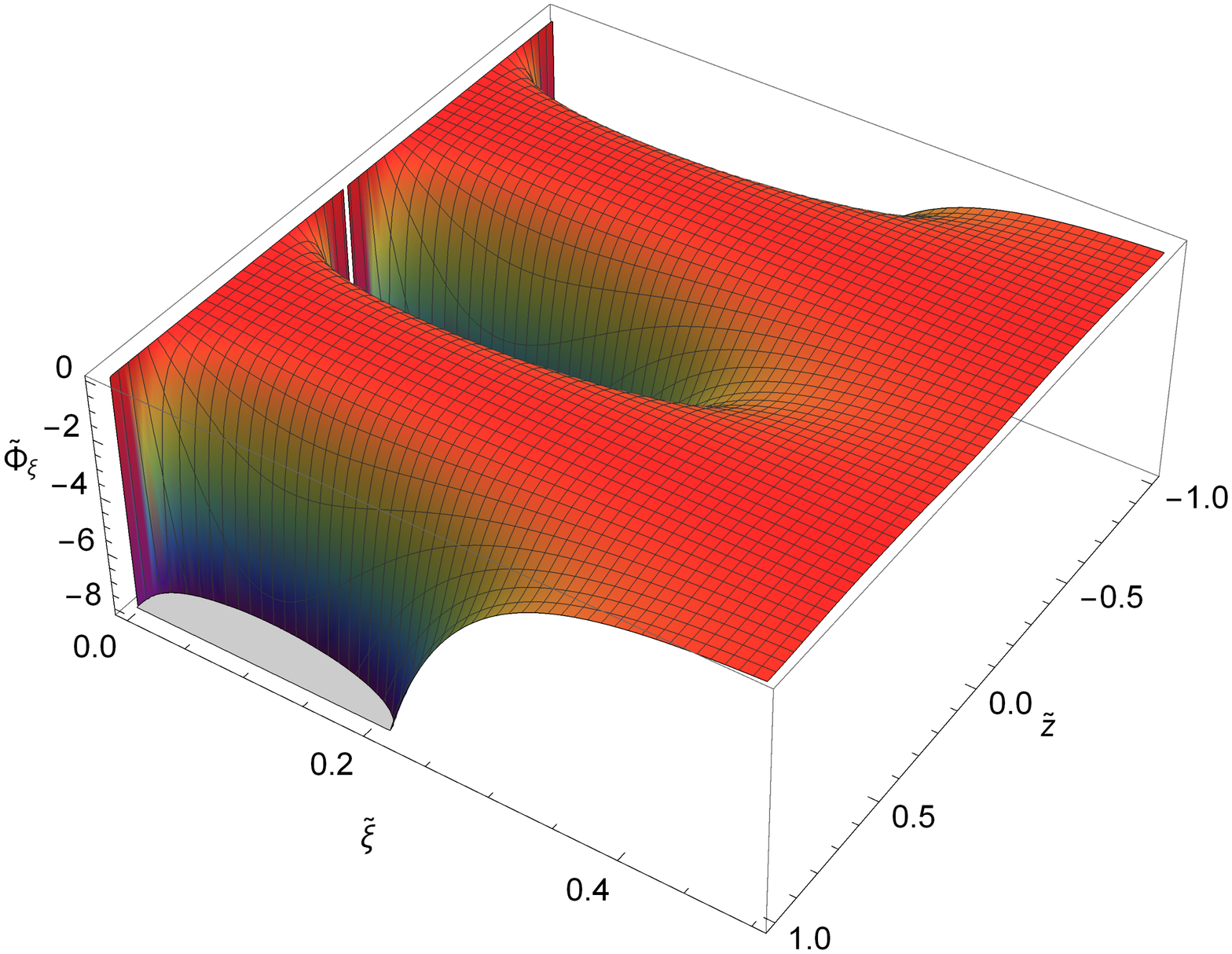}}
	\caption{Rescaled $\xi$-component of the gravitational force $\tilde{\Phi}_{\xi}\equiv \partial \tilde\Phi/\partial \xi$ in the cases $\tilde\lambda_{\mathrm{eff}} = 0.01$ and $\tilde\lambda_{\mathrm{eff}} = 0.1$ (left and right panels, respectively).}
	\label{fig:5}
%\end{figure*}
\vspace{0.6cm}
%\begin{figure*}
	\resizebox{0.48\textwidth}{!}{\includegraphics{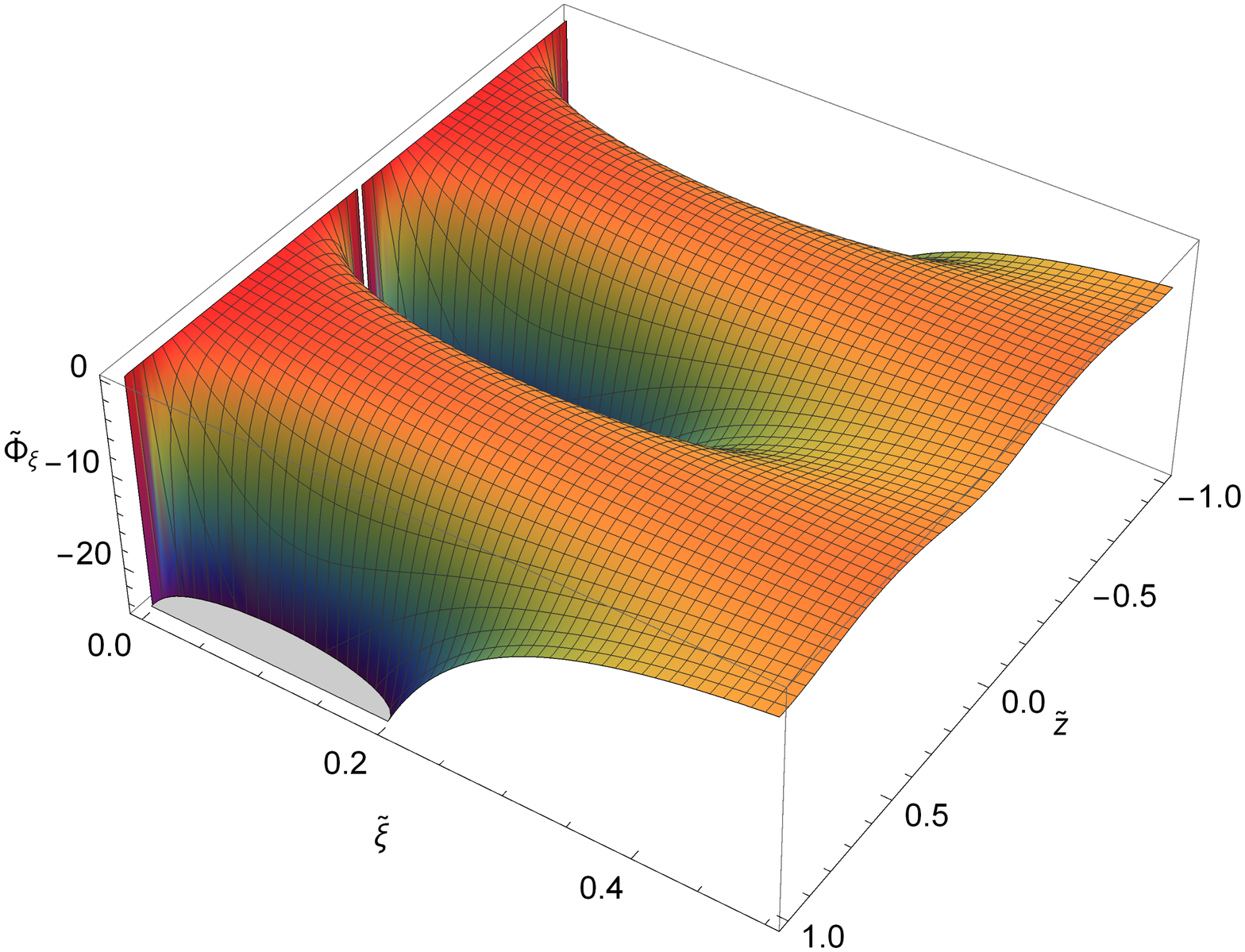}}\quad\quad
	\resizebox{0.48\textwidth}{!}{\includegraphics{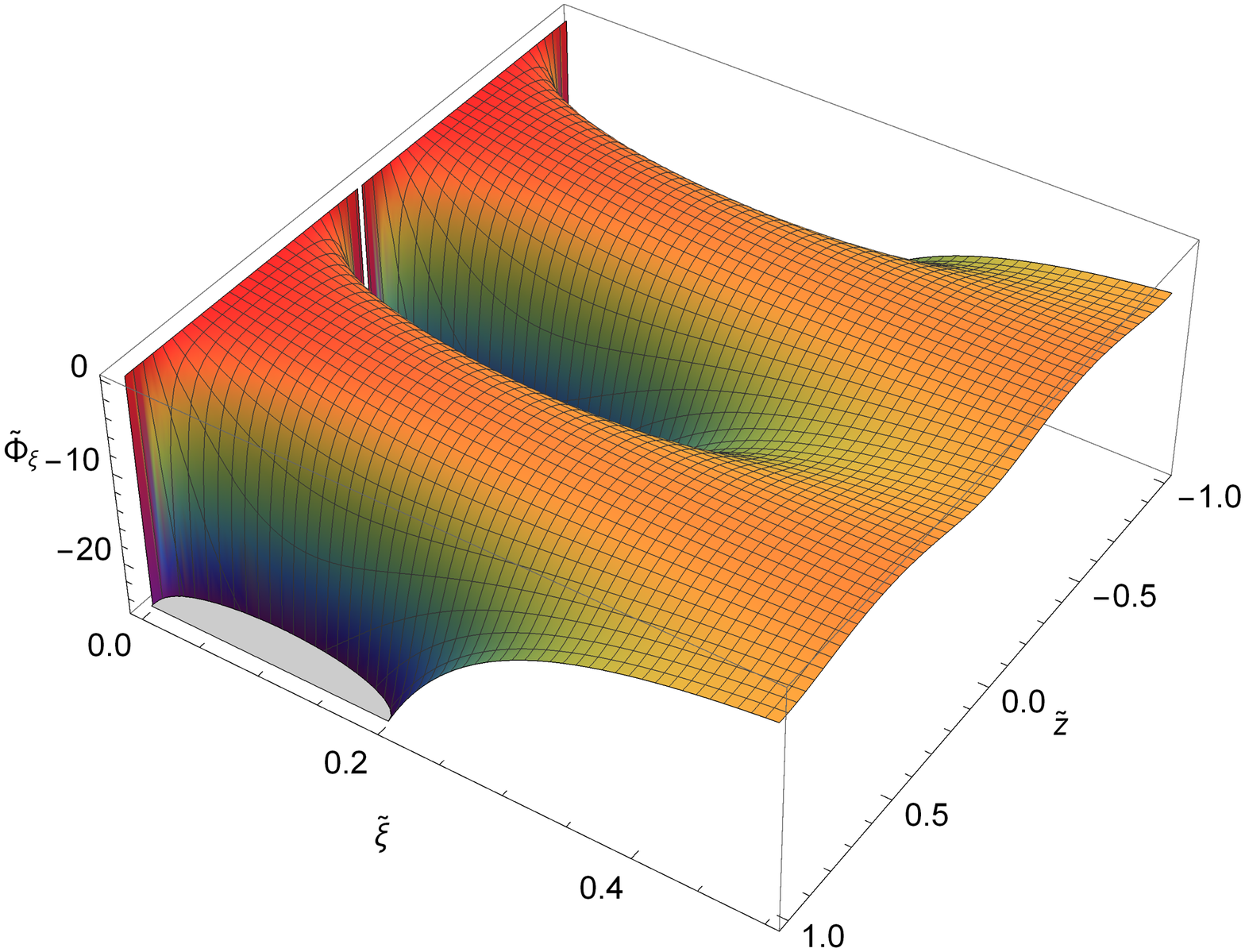}}
	\caption{Rescaled $\xi$-component of the gravitational force $\tilde{\Phi}_{\xi}\equiv \partial \tilde\Phi/\partial {\xi}$ in the cases $\tilde\lambda_{\mathrm{eff}} = 1$ and $\tilde\lambda_{\mathrm{eff}} = 10$ (left and right panels, respectively).}   
	\label{fig:6}
\end{figure*}

The behavior of the rescaled $\xi$-component of the gravitational force $\tilde{\Phi}_{\xi}$ for four chosen 
values of $\tilde\lambda_{\mathrm{eff}}$ is depicted (with the help of Mathematica \cite{Math}) in Figs.~\ref{fig:5},~\ref{fig:6}. To this end, we use the formula \rf{4.4} with  $n \gg n_{\exp}$ (where the numbers $n_{\exp}$ are given by Tables~\ref{results_table_5},~\ref{results_table_6}). 

%%%%%%%%%%%%%%%%%%%%%%%%%%%%

%%%%%%%%%%%%%%%%%%%%%%%%%%%%%%%%%%%%%%%%
%%%%%%%%%%%%%%%%%%%%%%%%%%%%%%%%%%%%%%%%%%%
%%%%%%%%%%%%%%%%%%%%%%%%%%%%%%%%%%%%%%%%%%%
%%%%%%%%%%%%%%%%%%%%%%%%%%%%%%%%%%%%%%%%%%

\section{Conclusion}\label{sec:5}

\setcounter{equation}{0}

In this paper we have studied the effect of the slab topology $T\times R\times  R$ of the Universe on the form of the  gravitational potential and force. We have found two alternative forms of the solution: one (see eq.~\rf{2.29}) is based on the Fourier series expansion of the delta function using the periodical property along the toroidal dimension, and another one (see eq.~\rf{2.30}) is obtained by direct summation of the solutions of the Helmholtz equation for a source particle and all its images. The latter solution takes the form of the sum of Yukawa-type potentials. For both of these alternative presentations, the screening length $\tilde\lambda_{\mathrm{eff}}$ is an important parameter. The physical meaning of this length can be most clearly seen from the second formula: it defines the distance (from the source or its image) at which the corresponding potential undergoes the exponential cutoff. According to the observations, $\tilde\lambda_{\mathrm{eff}} < 1$ for the present Universe. 

One of the main purposes of the paper was to determine which of the found alternative formulas works better from the point of numerical calculations. ``Better'' means which of the formulas requires less terms of the series to achieve the necessary precision. Our calculations show that for both gravitational potential and force, the formula with direct summation of Yukawa potentials is preferable in the physically relevant case $\tilde\lambda_{\mathrm{eff}} < 1$.  Additionally, in Figs.~\ref{fig:1}-\ref{fig:6}, we have presented graphically the gravitational potentials and force projections for four screening lengths $\tilde\lambda_{\mathrm{eff}}= 0.01, 0.1, 1, 10$, respectively.

\section*{Declarations}

{\noindent {\bf Funding.} The work of M. Eingorn, N. O'Briant and K. Arzu was supported by National Science Foundation (HRD Award \#1954454).
	
\ 

{\noindent {\bf Conflicts of interest.} The authors have no conflicts of interest to declare that are relevant to the content of this article.

\ 

{\noindent {\bf Data availability.} All data generated or analyzed during this study are included in this article.}

\ 

{\noindent {\bf Authors' contributions.}	
{\bf Maxim Eingorn:} Conceptualization, Methodology, Formal analysis, Investigation, Writing -- Review \& Editing, Visualization, Supervision, Project administration, Funding acquisition. {\bf Niah O'Briant:} Formal analysis, Investigation, Visualization. {\bf Katie Arzu:} Formal analysis, Investigation, Visualization. {\bf Maxim Brilenkov:} Formal analysis, Investigation. {\bf Alexander Zhuk:} Methodology, Formal analysis, Investigation, Writing -- Original Draft, Supervision.
}

\end{document}